\documentclass{PoS}

\usepackage{cite}

\newcommand{\eq}{\begin{equation}}
\newcommand{\eqx}{\end{equation}}
\newcommand{\eqn}{\begin{eqnarray}}
\newcommand{\eqnx}{\end{eqnarray}}
\newcommand{\bi}{\begin{itemize}}
\newcommand{\ei}{\end{itemize}}
\newcommand{\nn}{\nonumber}
\newcommand{\ra}{\rangle}
\newcommand{\la}{\langle}
\newcommand{\bz}{\bar{z}}

\title{Beyond Complex Langevin Equations: a Progress Report}

\ShortTitle{Beyond Complex ...}

\author{\speaker{Jacek Wosiek}\\
        Jagiellonian University, Cracow, Poland\\
        E-mail: \email{Jacek.Wosiek@uj.edu.pl}}

\author{Blazej Ruba\\
        Jagiellonian University, Cracow, Poland\\
        E-mail: \email{Blazej.Ruba@doctoral.uj.edu.pl}}

\abstract{After a short review of one of proposals to avoid complex stochastic processes in Complex Langevin studies, the recent progress in the former is reported.
In particular, the new developments allow now to construct positive and normalizable representations for gaussian quantum mechanical, as well as field theoretical, path integrals directly in the Minkowski time. A relation to the idea of thimbles  is also discussed.
}

\FullConference{The 36th Annual International Symposium on Lattice Field Theory - LATTICE2018\\
		22-28 July, 2018\\
		Michigan State University, East Lansing, Michigan, USA.}

\begin{document}

\section{Introduction and reminder}
The well known difficulties with the Complex Langevin method \cite{P,Kl,AY,HW,S4,S2,S1,NS,ES} resulted recently in the "Beyond Complex Langevin" approaches \cite{Sa,We,Sa2,SW,Ja}. They do not use the problematic complex random walks to generate equivalent probability distributions $P(x,y)$. Instead, one attempts  to construct $P(x,y)$ directly from the matching equations
\eqn
\la f(x) \ra_{\rho(x)}=\la f(x+i y) \ra_{P(x,y)}, \label{B1}
\eqnx
which are in fact the only relevant requirements. 
We briefly remind here the idea of \cite{Ja} and then report on its recent developments.

The trick was to introduce a second, {\em independent} complex variable $\bar{z}$, and rewrite Eq. (\ref{B1})  as (all contours are chosen such that the integrals exist)
\eqn
 \frac{\int_{\Gamma_z} f(z) \rho(z) dz}{\int_{\Gamma_z} \rho(z) dz} = \frac{\int_{\Gamma_z} \int_{\Gamma_{\bz}} f(z) P(z,\bz) dz d\bz }{\int_{\Gamma_z} \int_{\Gamma_{\bz}} P(z,\bz) dz d\bz}. \label{B3}
\eqnx
In addition, the positivity and normalizability of $P(z,z^*)=P(x,y)$ was required.
The first condition is simple to achieve. Since an observable $f(z)$ is independent of $\bz$, any $P$ which satisfies 
\eqn
\rho(z)=\int_{\Gamma_{\bz}} P(z,\bar{z}) d \bar{z},  \label{Pro}
\eqnx
will solve (\ref{B3}). However assuring the second part is not straightforward.
Nevertheless this program could in fact be carried through in some interesting cases as discussed next.

\noindent {\em The generalized gaussian model}

In the first example the well known solution of Ambjorn and Yang \cite{AY}, for the complex gaussian wieght $\rho(x)=\exp{\left(-\sigma x^2/2\right)}$ with a complex slope $\sigma\in{\cal C}$.  Present prescription allows to rederive it in a  
 slightly more general form. To assure positivity, we begin with a generic quadratic action for $P(z,\bar{z})$ in two complex variables $z$ and $\bz$
\eqn
P(z,\bar{z})=e^{-S(z,\bar{z})},\;\;\;\;S(z,\bar{z})&=& a^* z^2 + 2 b z \bz + a \bz^2, \label{S2}
\eqnx
with an arbitrary complex $a=\alpha+i\beta$ and real $b=b^*$. Indeed, for $\bar{z}=z^*$,  
$P(x,y)$ is positive and normalizable provided $b > |a|$. On the other hand, the the $\bar{z}$ integral 
\eqn
\rho(z)=\int_{\Gamma_{\bz}} P(z,\bz) d \bz = 
  \frac{1}{2}\sqrt{\frac{\pi}{-a}}\exp{\left( - s z^2\right)},\;\;\;s=\frac{|a|^2-b^2}{a}. \label{ip2}
\eqnx
reproduces the original complex action. Therefore Eq.(\ref{S2}) gives directly the solution of  \cite{AY} when restricted to $\bz=z^*$. Moreover, it also provides its generalization for all complex values of the slope $s$ in (\ref{ip2}).
This is best seen in two special cases below.

1.   The complex density blows up along the real axis for real and negative $s$. Yet the distribution $P(x,y)$ is positive, normalizable at $\beta=0$, $0<\alpha< b$, and reproduces the correct analytical continuation of the average over the "divergent" distribution $\rho$. This accounts for a ``striking example" observed in the literature. Namely, it was found that upon change of variables, the complex Langevin simulation based on (\ref{ip2})  actually has the correct fixed point also for negative ${\cal R}e\;s$. It is seen now that the positive distribution used in this case is a part of a richer structure (\ref{ip2}), which accommodates negative ${\cal R}e\;s$ as well.
 
2. Similarly, the complex density $\rho(z)$ for purely imaginary $s$ is readily represented by the positive distribution $P(x,y)$, which is perfectly well defined at $\alpha=0$ and $\beta$ with $|\beta|<b$. Interestingly this allows to construct positive representations for Feynman path integrals directly in the Minkowski time, see below.

\noindent {\em A nonlinear model }

 Another equivalent, positive distribution was derived starting from the action
\eqn
S_4(z,\bz)=(a^* z^2+2 b z \bz + a \bz^2)^2, \nn  
\eqnx
with complex $a$  and real $b$. The density $P_4(x,y)$ is again positive and normalizable on the $x,y$ plane.
The complex density $\rho_4(z)$ can then be readily obtained as
\eqn
\rho_4(z)&=&\frac{i}{2}\int_{\Gamma_{\bz}} d \bz  e^{- S_4(z,\bz)}
=\frac{i}{2}\left(\frac{1}{2 a^2}\right)^\frac{1}{4}  \exp{\left(-\sigma z^4\right) } \left(\sigma z^4 \right)^\frac{1}{4} 
K_{\frac{1}{4}}\left(\sigma z^4\right),\label{cw4}
\eqnx
with an arbitrary complex
\eqn
\sigma=\frac{ (b^2-|a|^2)^2}{2 a^2}.\nn .
\eqnx
Again all contours are such that the integrals exists. Essentially one can choose straight lines with appropriate slopes determined by 
the phase of $a$. For the first two nonvanishing moments matching conditions were checked by explicit calculations. Hence we believe that our trick is more general, and indicates possible existence of some unexplored yet structures.

\noindent {\em Many variables and quantum mechanical Minkowski path integrals}
%===============================================

A generalization to arbitrary number of gaussian variables, hence also to quantum mechanical path integrals was done as well. Feynman path integrals in Minkowski time 
correspond to purely imaginary slope $s$ in (\ref{ip2}), for which our approach is perfectly well applicable. Consequently the working prescription to 
calculate Minkowski path integrals by stochastic, i.e., Monte Carlo methods is currently available. 

The action consists now of $N$ copies of (\ref{S2})  with additional nearest neighbour couplings. Periodic boundary conditions are assumed: $z_{N+1}=z_1, \bz_{N+1}=\bz_1, z_{0}=z_N, \bz_{0}=\bz_N$, $a, c \in C$, $b \in R$.
\eqn
S_N(z,\bz)= \sum_{i=1}^N
 a \bz_i^2 + 2 b  \bz_i z_i + 2 c \bz_i z_{i+1} + 2 c ^* z_i \bz_{i+1} + a^* z_i^2,\;\;\;\;P_N(z,\bz)=e^{-S_N(z,\bz)} . \label{SN} 
\eqnx
Integrating $P_N(z,\bz)$ over all $\bz$ variables gives the complex density 
\eqn
\rho(z)=\int \prod_{i=1}^{N}d\bz_i P(z,\bz)=\left(\frac{i}{2}\right)^N\int \prod_{i=1}^{N}d\bz_i \exp{\left(-S_N(z,\bz)\right)}\equiv
\exp\{-S^{\rho}_N\}.\nn
\eqnx
One obtains for the effective action ( $2c=2\gamma=-b + |a|$), 
\eqn
-S_N^{\rho}(z)= {\cal A} \sum_{i=1}^N    \left(   z_i -  z_{i+1} \right)^2
- r\left(z_{i-1}-z_{i+1}\right)^2,\;\;\;\;{\cal A}=\frac{b(b-|a|)}{a} ,\;\;r=\frac{b-|a|}{4b}.\;\;\;\;\label{Srho}
\eqnx
This is similar, but not identical, to the discretized Feynman action for a free particle. 
\eqn
S^{free}_N=\frac{i m}{2 \hbar\epsilon} \sum_{i=1}^N (z_{i+1}-z_i)^2.  \label{SFd}
\eqnx 
However, both constraints, namely ${\cal A}  \rightarrow \frac{i m}{2 \hbar\epsilon}$ ,  and $r= \rightarrow 0$,  
 can be satisfied in the limit  (referred from now on as $\lim_1$)
\eqn
|a|, b \rightarrow \infty, b-|a|=\frac{ m}{2 \hbar\epsilon}=const.\equiv d,\;\;a=-i |a|\equiv i\beta . \label{lim1}
\eqnx
All quantum averages can now be obtained by weighting suitable, i.e. complex in general, observables with the positive and normalizable distribution $P_N(x_i,y_i)=\exp{\left(-S_N(z_i,z_i^*)\right)}$
, and then taking  the limit (\ref{lim1}) followed by the continuum limit: $N\rightarrow\infty,\epsilon\rightarrow 0, N\epsilon=const\equiv T $. 
This concludes the construction of the positive representation for the path integral of a free particle directly in the Minkowski time.

With somewhat different identification of parameters of the action (\ref{SN}) one covers as well the harmonic oscillator case (with the frequency $\omega$). 
\eqn
-S_N^{\rho}(z)=\frac{i m }{2 \hbar \epsilon}\left( (z_1-z_2)^2 - \frac{\omega^2 \epsilon^2}{2}(z_1^2+z_2^2)\right)+ (nnn) .  \nn
\eqnx
As before, the $nnn$ terms vanish for large $|a|$ and $b$. The first limit ($lim_1$) is now taken 
along a different trajectory. In terms of one independent parameter $\nu\rightarrow 0$, it reads ($a=-i |a|$)
\eqn
b=\frac{\mu}{\nu},\;\;\; |a|= \frac{\mu}{\nu} \zeta(\nu,\rho),\;\;\; 2\gamma=-\mu \zeta(\nu,\rho),  \;\;\;
\zeta(\nu,\rho)= \frac{\sqrt{1-2\nu^2\rho+\nu^2\rho^2}-\nu(1-\rho)}{1-\nu^2},   \label{hotraj}
\eqnx
with $\mu$ and $\rho$ depending on $N$ and on parameters of the harmonic oscillator in the continuum
\eqn
\rho=\frac{\omega^2 T^2}{2 (N-1)^2},\;\;\;
\mu=\frac{m (N-1)}{2\hbar T}.   \nn
\eqnx
This is the main modification compared to a free particle. With the first limit taken along the trajectory (\ref{hotraj}) the action (\ref{SN}) provides a positive representation for Minkowski path integral of a one-dimensional harmonic oscillator. 

The restricted $P(z_i,z_i^*)$ is positive as evident from the construction (\ref{SN}) while the normalizability follows from inspecting eigenvalues of the real form $S_N({x_i,y_i})$. All but one of them are indeed positive.  In a free particle case, due to the translational symmetry, one eigenvalue is zero. It becomes negative for the harmonic oscillator representation. One can deal with them in the standard manner.

\begin{figure}[h]
\begin{center}
\includegraphics[width=7cm]{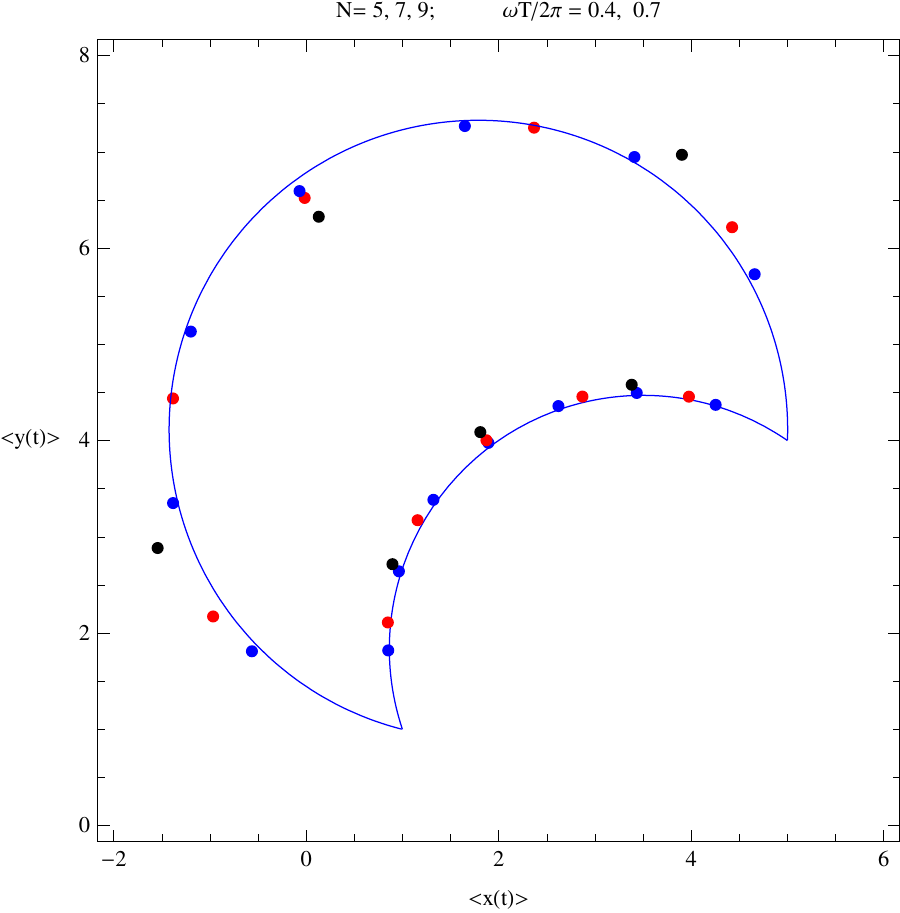}
\end{center}
\vskip-4mm \caption{ The average position of a quantum particle in a constant magnetic field calculated with two different bounadry conditions. Points represent the averages over positive distribution, described in the text, after performing the first limit. Solid lines show exact results.} \label{fig1}
\end{figure}

The most interesting application is perhaps that of a charged particle in a constant magnetic field. The standard formulation does not have a positive representation
even after the Wick rotation. Yet the present approach provides a positive solution. It is well known since the time of Landau that the problem can be mapped into a one dimensional harmonic oscillator with the shifted centre of oscillations. Hence we need only to identify appropriate parameters. A sample of results is shown in Fig.\ref{fig1}.
\section{New developments}%================================================================
During the last year our theoretical understanding of the whole approach was improved which resulted in further interesting applications \cite{BR}.  Here I will only list some of them and point out their relevance. For more detailed discussion 
see the following talk by Blazej Ruba .

{\em Multidimensional Cauchy equation} The role of the $\bz$ variable in the above analysis, Eq.(\ref{B3}) and the following, looks somewhat ambiguous at first sight and requires careful analysis. First $\bz$ is 
considered as independent of $z$, however later it is set to be equal to complex conjugate of $z$. In particular equivalence of (\ref{B1}) and (\ref{B3}) requires, first  
complexifying x and y,  and then proving the following relation (${\cal R}_{x(y)}$ denotes the real axis in the complex $x(y)$ plane)
\eqn
 \int_{\Gamma_z} \int_{\Gamma_{\bz}} f(z) P(z,\bz) dz d\bz =  \int_{{\cal R}_x} \int_{{\cal R}_y}  f(x+iy) P(x,y) dx dy \nn
\eqnx
which connects two double-contour integrals in the four-dimensional space ${\cal C}^2$. This was done with the aid of the "four-dimensional" Cauchy equation.

{\em A family of solutions of the gaussian toy model and emergence of thimbles}  It is well known that there exist a one parameter family of positive distributions equivalent to the 
complex gaussian weight \cite{AY,Ja}. However Ruba was able to exemplify this freedom in a rather elegant way. Namely all solutions have a form (\ref{ip2}) with
\eqn
a=-\mu\sigma^*,\;\;\;\; b=|\sigma|\sqrt{\mu(\mu+1)}
\eqnx
with a free parameter $0 < \mu < \infty$ labelling each case. There are two interesting special situations.
\begin{equation}
\mu \longrightarrow \left\{ \begin{array}{ll}
                 \left(\frac{Re \sigma}{Im \sigma}\right)^2 ,\;\;\;\;\;  &  P(x,y) \Rightarrow AY \\  &  \nn  \\   \nn 
                                                                          \infty, \;\;\;\;\; & P(x,y) \rightarrow \delta(Im \sqrt{\sigma} z ) \exp{(-\sigma z^2)}   \nn
                                      \end{array}  \nn
      \right. \nn
\end{equation}
Namely our solution reproduces the Ambjorn Yang result for a particular value of $\mu$ and, even more interestingly, at infinite $\mu$ it is concentrated along the line of real $
\sqrt{\sigma} z$. This line is nothing but a linear thimble in this simple case. Hence the connection between a complex Langevin dynamics and thimble construction has been readily
established in this case.

{\em Toy model - a quartic action} The analysis of the toy model with the quartic action has also been extended. In particular all moments have been calculated analytically in both ways, i.e. as the "averages"  over the complex weight (\ref{cw4}) and reproduced using the statistical average over $P_4(x,y)=\exp{\left(-S_4(z,z^*)\right)}$. This completes the proof of equivalence of both distributions and shows that the method can be applied to the non-linear cases as well. Again there exists a one-parameter family of equivalent positive distributions. Moreover, thimbles emerge again in a limit analogous to the previous case. The analysis is more involved, but indeed the two-dimensional positive density becomes concentrated on a one-dimensional, linear support.

{\em Path integrals} Generalization of above toy model to  quantum mechanical path integrals have also been improved and augmented with new, interesting results.

\begin{figure}[h]
\begin{center}
\includegraphics[width=7.5cm]{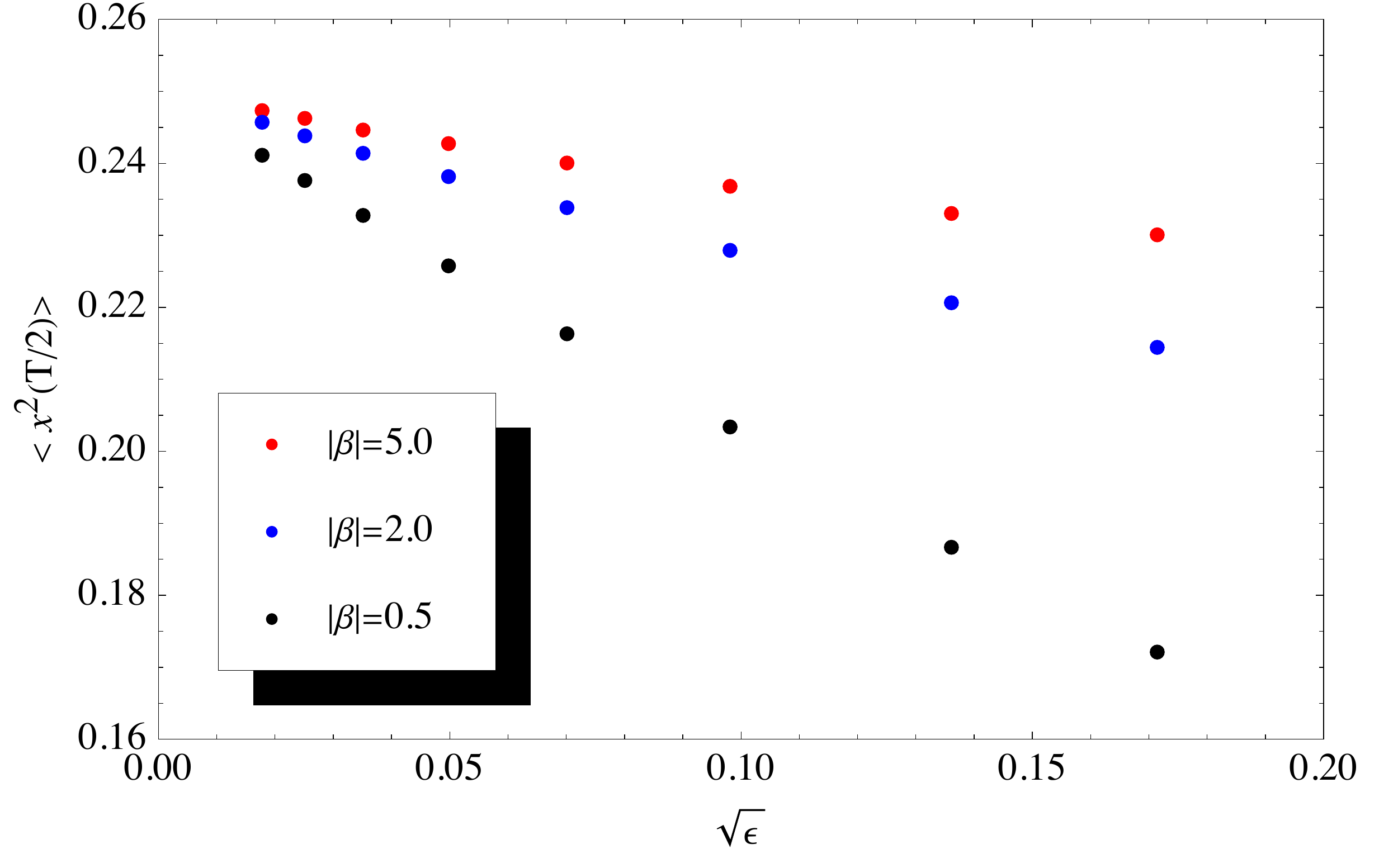}
\includegraphics[width=7.5cm]{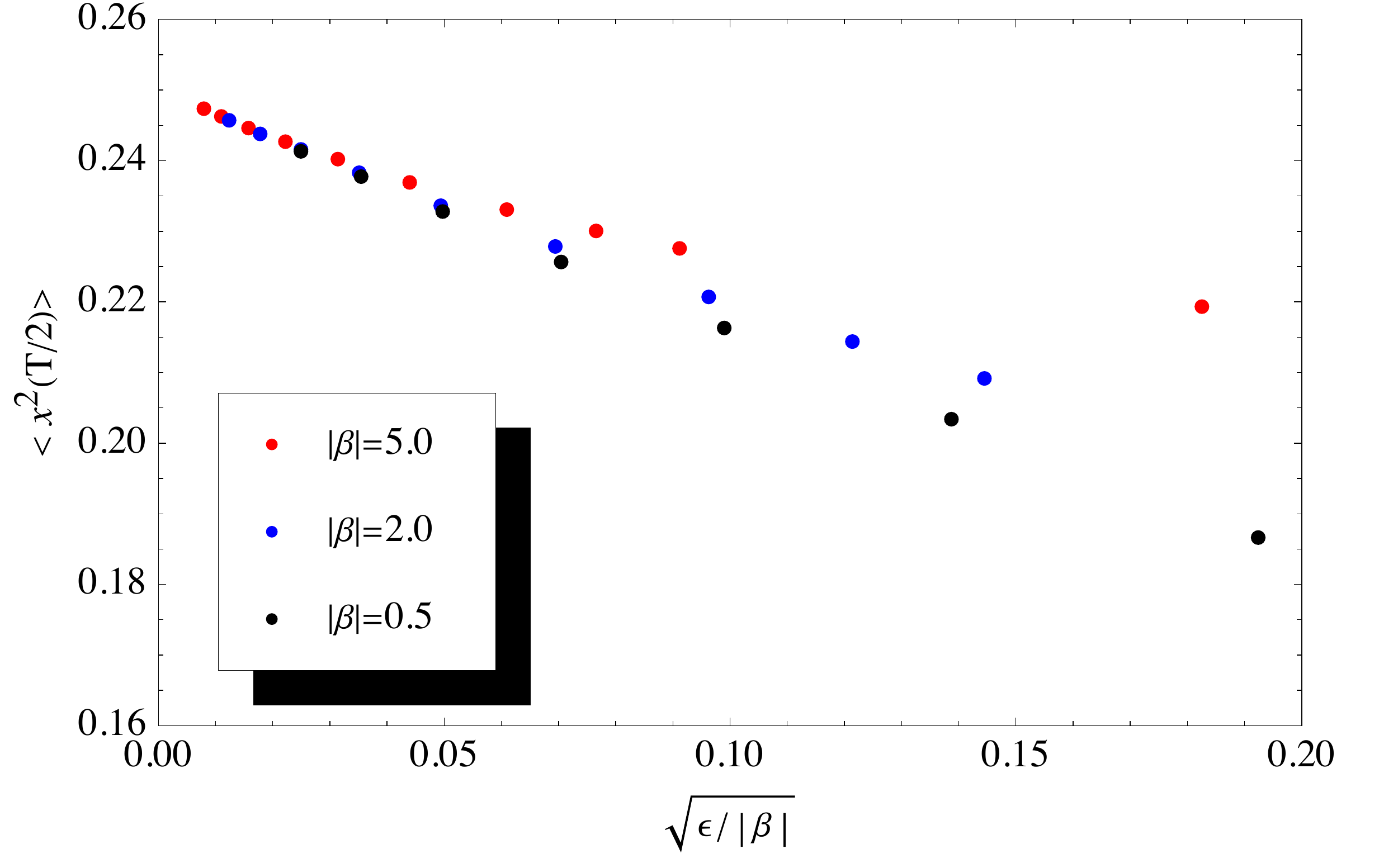}
\end{center}
\vskip-4mm \caption{Left panel - a measure of a dispersion of a free quantum particle as a function of the discretization scale, $\epsilon$, for different $\beta$ (c.f. (\ref{lim1}). Right panel - the same quantity vs., the scaling variable $\epsilon/|\beta|$.} \label{fig2}
\end{figure}

{\em A free particle} As explained above the effective action, which results from the the integrals over $\bz's$, contains an unwanted next-to-nearest-neighbour terms, c.f.(\ref{Srho}). Therefore, before taking the continuum limit, the additional limit (referred above as $lim_1$) had to be performed to get rid of these couplings.  
Studying a particular two point correlation it was shown \cite{BR} that in fact one can interchange both limits, i.e. the continuum limit can be taken first without changing the final answer. This is illustrated in Fig.\ref{fig2} for yet another observable (an absolute value of a dispersion  in a middle of a free Minkowski quantum trajectory, with the end points fixed at the origin). 

It is readily seen (left panel) that different orderings of both limits ($\beta\rightarrow\infty$ and $\epsilon\rightarrow 0$ give the same exact result of 1/4. Moreover the analytic analysis in \cite{BR} shows that the only relevant, i.e. scaling, variable is  $\epsilon/|\beta|$ which is confirmed in panel 2 of the Figure.

{\em  Thimbles in many variables} Interestingly Ruba has also found thimbles in this case. It turns out they have a very simple geometry: all contours should be rotated by 45 degrees.

{\em Harmonic oscillator - the origin of negative eigenvalues and a new solution of the problem} For a fee particle all, but one, eigenvalues of the real action $S(\vec{x},\vec{y})$ are positive giving rise to the well behaved probability, except along one one-dimensional valley which results from the translational symmetry. In the harmonic oscillator case this eigenvalue became negative and in principle hampered our criterion of normalizability.
Ruba has pointed out that this negative eigenvalue is the consequence of the Morse theorem, and in fact there may be more of them depending on how many focal points are crossed by classical trajectory during the evolution time $T$. 

Moreover, he also proposed a simple solution to avoid this problem, namely by rotating corresponding contours by 45 degrees, similarly to the free particle case. As a consequence the Minkowski correlation function could (and was) calculated by MC methods for the first time (see the following talk for the results).
 
{\em Free Minkowski scalar field theory in any dimensions} Generalization to the field theory is straightforward. There are more negative eigenvalues, corresponding to larger number of oscillators, but their origin is again the same. Similarly as in quantum mechanical path integrals, Ruba construction of positive and normalizable densities, equivalent to Minkowski weights, works. The only bound comes from the usual increase of the number of degrees of freedom, but it is by no means prohibitive. As an example he calculated {\em by Monte Carlo} the standard Minkowski propagator, in two dimensions, revealing nicely its causal structure (c.f. the next talk).
\section{Summary}
Instead of using problematic complex random walks, Beyond Complex Langevin approaches attempt to construct pairs of corresponding (i.e. complex vs. probabilistic) weights directly from the matching equations for observables. The method reviewed here consists of extending the complex action to the two independent complex variables, thereby simplifying the relation between analytic continuations of above weights. In the second step one constructs solutions which restricted to real arguments are positive and normalizable.  

The trick works fine for the one dimensional gaussian, and a specific quartic, toy models. In particular the well known solution of the gaussian model was extended to arbitrary complex slopes.

The technique was then generalized to many gaussian variables, including path integrals with purely imaginary slopes. That is, the positive, probabilistic representation of Minkowski gaussian path integrals was found without performing the Wick rotation. The method was then applied to some textbook quantum mechanical examples. The most interesting one is the motion of a quantum particle in an external, constant magnetic field, since it did not admit, until now, a positive representation, even in the Euclidean time.

Recently, our theoretical understanding of the main principle was augmented by more formal arguments. The analysis of the nonlinear toy model was also fully completed showing that indeed the approach works in some nonlinear cases. Interesting relation between the Langevin dynamics and existence of thimbles was also found.

There is also a progress in applications to path integrals. The new, double limit was proposed which reproduces the continuum physics under weaker constraints. In particular an interesting scaling variable was identified. Again occurrence of thimbles (this time in the functional space) was explicitly demonstrated.

A negative eingenvalue, which appeared in the harmonic oscillator case was understood in terms of the Morse theorem and a simple remedy of this problem was proposed. As a consequence the well defined algorithm to simulate Minkowski path integral was designed and the real time correlation function was calculated by Monte Carlo simulations.

Finally, the method generalizes straightforwardly to free field theories in arbitrary dimensions. As an illustration the causal structure of the two-dimensional Minkowski scalar propagator was reproduced by well behaved  Monte Carlo simulations.

A range of interesting applications is conceivable. One possibility is Monte Carlo study of the real time evolution, i.e. also particle production, 
in external fields \cite{FG}.

\vspace*{.5cm}

We thank Francois Gelis for instructive discussions. JW thanks Michael Ogilvie for interesting discussion and comments.

This work is supported in part by the NCN grant: UMO-2016/21/B/ST2/01492.

\end{document}